# Text to Speech Synthesis

Harini S
Assistant Professor
Department of Information Science and Engineering
BMS College Of Engineering
Bengaluru, India
harinis.ise@bmsce.ac.in

Manoj G M
Department of Information Science and Engineering
BMS College Of Engineering,
Bengaluru, India
manojgm.is21@bmace.ac.in

*Abstract* - Text-to-Speech (TTS) synthesis is a technology that converts written text into spoken language, allowing for the generation of natural-sounding speech. This process involves various stages, including text analysis, linguistic processing, and waveform synthesis. TTS systems use algorithms and models to transform input text into corresponding speech signals, mimicking the intonation, rhythm, and prosody of human speech. Over the years, TTS synthesis has evolved, incorporating neural network-based approaches to enhance the naturalness and expressiveness of generated speech. Applications of TTS span across accessibility tools, voice assistants, navigation systems, and entertainment, offering a versatile and impactful solution for converting written content into audible form. Ongoing research continues to refine TTS systems, striving to achieve even greater levels of realism and adaptability in generating human-like speech.

## I. INTRODUCTION

Text-to-Speech (TTS) technology has revolutionized the way we interact with computers and devices by enabling them to convert written text into spoken words. TTS systems utilize advanced linguistic algorithms and artificial intelligence techniques to generate high-quality, natural-sounding speech that closely resembles human speech patterns and intonations. the primary goal of TTS is to provide an inclusive and accessible means of communication for individuals with visual impairments or those who prefer auditory information. By converting written text into speech, TTS technology allows people to listen to text-based content such as books, articles, web pages, and messages, enhancing their overall digital experience and facilitating information consumption.

TTS systems have come a long way in recent years, thanks to advancements in deep learning, neural networks, and data-driven approaches. Modern TTS models are trained on vast amounts of multilingual and multi-speaker data, enabling them to produce speech in different languages and mimic various voices with remarkable accuracy and expressiveness. moreover, TTS has found applications beyond accessibility, benefiting industries like entertainment, education, customer service, and more. It enables voice assistants, virtual characters, and automated systems to communicate with users, enhancing user experience and making human-machine interactions more intuitive and engaging. as TTS technology continues to evolve, we can expect further improvements in speech quality, customization options, and efficiency. These advancements will contribute to creating more immersive and inclusive experiences, where technology seamlessly integrates with our daily lives, offering an enhanced way to consume and interact with information. TTS systems often provide options for personalization and customization. Users can choose different voices, accents, and speech rates according to their preferences. This flexibility allows individuals to tailor the synthesized speech to their liking, enhancing their overall experience and engagement with the content.

## II. PROBLEM STATEMENT

Text-to-speech synthesis is like teaching computers to talk. The main problem is making the computer-generated speech sound as natural as possible, like a human speaking. This involves dealing with different languages, making the voice express emotions, and responding quickly. It's also important to let users choose how the computer voice sounds and to make sure it works well for people with disabilities. In simpler terms, the challenge is to make the computer talk in a way that sounds real and is easy for everyone to understand. Handling different languages and accents, expressing emotions, working in real-time, letting users customize the voice, dealing with lots of words, and being useful for people with disabilities. Solving these challenges involves using smart computer programs and algorithms that understand language and sound to create better and more human-like speech from written text.

## III. LITERATURE SURVEY

[1] Text to Speech Conversion based on Emotion using Recurrent Neural Network.
The paper introduces an enhanced Text-to-Speech Conversion System (ETTS) incorporating emotions through Recurrent Neural Network (RNN) technology. Four fundamental emotions, including 'happy,' 'sad,' 'anger,' and 'neutral,' are detected with high accuracy using the GRU model within the RNN. The dataset is created by combining existing datasets like dailydialog, emotion-stimulus, and isear. The model architecture comprises layers such as Embedded Layer, Bidirectional GRU Layer, and Dense

Layers. Achieving an accuracy of 86.77% on a combined dataset, the system has promising applications in storytelling and aiding visually impaired individuals. Future improvements may include standalone applications, additional emotions, and background music integration.

[2] Neural Speech Synthesis with Transformer Network.
This paper aiming to synthesize intelligible and natural audios which are indistinguishable from human recordings. Traditional TTS systems have two components: front-end and back-end. Front-end is responsible for text analysis and linguistic feature extraction, such as word segmentation, part of speech tagging, multi-word disambiguation and prosodic structure prediction; back-end is built for speech synthesis based on linguistic features from front-end, such as speech acoustic parameter modeling, prosody modeling and speech generation. In the past decades, concatenative and parametric speech synthesis systems were mainstream techniques. However, both of them have complex pipelines, and defining good linguistic features is often time-consuming and language specific, which requires a lot of resource and manpower. Besides, synthesized audios often have glitches or instability in prosody and pronunciation compared to human speech, and thus sound unnatural.

[3] Glow-TTS: A Generative Flow for Text-to-Speech via Monotonic Alignment Search.
This paper aim to introduces a novel text-to-speech (TTS) synthesis model called Glow-TTS. This model is designed to generate mel-spectrograms conditioned on a monotonic and non-skipping alignment between text and speech representations. Glow-TTS combines the best of both worlds in terms of robustness, diversity, and controllability by employing both hard monotonic alignments and generative flows. The paper presents the training and inference procedures of Glow-TTS, as well as an alignment search algorithm that removes the necessity of external aligners from training. Additionally, it covers the architecture of all components of Glow-TTS, including the text encoder, duration predictor, and flow-based decoder. Glow-TTS is positioned as a diverse, robust, and fast TTS synthesis model with potential applications in AI voice assistant services, audiobook services, advertisements, automotive navigation systems, and automated answering services. However, the paper also acknowledges potential concerns related to the abuse of TTS models for cyber crimes and the need for careful usage in critical domains such as news broadcast. The model's performance and capabilities make it a significant contribution to the field of text-to-speech synthesis, offering a promising solution for natural and controllable speech synthesis.

[4] Grad-TTS: A Diffusion Probabilistic Model for Text-to-Speech. A novel approach to text-to-speech synthesis by introducing the Grad-TTS model, which employs a diffusion probabilistic model. In the realm of text-to-speech (TTS), diffusion models are likely utilized to depict the gradual evolution of a system over time, possibly describing the step-by-step generation process of speech waveforms. The architecture of Grad-TTS would be a key focus, encompassing the intricacies of the diffusion-based probabilistic model. Training the model would involve a dataset comprising linguistic features (text) and corresponding acoustic features (speech waveforms), with the paper delving into the specifics of the training procedure, potentially involving gradient-based optimization methods. Evaluation metrics, both subjective (such as human listener assessments) and objective (related to speech quality), would likely be discussed to measure the performance of Grad-TTS.

[5] NaturalSpeech: End-to-End Text to Speech Synthesis with Human-Level Quality. A focus on achieving high-quality text-to-speech (TTS) synthesis through an end-to-end approach, possibly with a model named NaturalSpeech. The paper likely explores a comprehensive system that takes input text and directly generates human-like speech, eliminating the need for intermediate steps. The emphasis is likely on achieving a level of quality comparable to human speech. The architecture and methodology of the NaturalSpeech model would be central to the paper, detailing how the model processes and transforms text into natural-sounding speech. Training data composition and the training procedure would also be crucial aspects, with potential consideration for large datasets to capture diverse linguistic patterns. The evaluation section may include both subjective assessments, where human listeners rate the quality of synthesized speech, and objective measures assessing aspects like clarity, prosody, and naturalness.

[6] Flowtron: An Autoregressive Flow-based Generative Network for Text-to-Speech Synthesis. To introduces the Flowtron model, which is designed for text-to-speech synthesis. Utilizing an autoregressive flow-based generative network, the model is likely structured to generate speech waveforms in a step-by-step manner, capturing intricate dependencies in the input text. This approach may allow for the production of natural-sounding speech by considering the sequential nature of language. The paper likely elaborates on the architecture and training procedure of Flowtron, discussing how it processes text inputs and transforms them into high-quality speech outputs. Evaluation metrics, both subjective and objective, are probably presented to assess the model's performance in terms of naturalness and quality.
Flow-based architecture: It employs invertible transformations to learn a flexible latent space, enabling precise control over speech output and efficient likelihood-based training.
Autoregressive generation: It produces mel spectrograms one frame at a time, ensuring temporal coherence and natural-sounding speech.
Latent space manipulation: It allows for fine-grained control over speech characteristics like pitch, tone, speech rate, cadence, and even speaker style, through manipulation of the latent space.

[7] FastSpeech: Fast, Robust, and Controllable Text to Speech. It involves proposing a novel TTS model called FastSpeech. FastSpeech employs a non-autoregressive approach, departing from traditional autoregressive methods, to enhance speed in generating speech. It utilizes a sequence-to-sequence architecture with a feed-forward Transformer, enabling parallelization for faster synthesis. Additionally, the model introduces a pitch prediction module for better prosody control. the challenge of synthesizing high-quality and natural-sounding speech from text in a fast, robust, and controllable manner. Traditional Text-to-Speech (TTS) systems often face issues related to speed, robustness, and control over speech synthesis, and this work aims to improve these aspects. the effectiveness of FastSpeech in terms of speed, robustness, and controllability in comparison to traditional autoregressive TTS models. The model achieves competitive or superior performance in terms of

speech quality while significantly reducing synthesis time. The controllability aspect refers to the ability to manipulate speech attributes, such as pitch, with improved accuracy. The results are often evaluated using objective metrics and subjective evaluations, such as Mean Opinion Score (MOS), to assess the quality of synthesized speech.

[8] Text – To – Speech Synthesis (TTS). aimed at converting written text into spoken language. The primary goal is to facilitate quick assimilation of textual information and support reading development. The system targets individuals with physical impairments, including the deaf, dumb, blind, and elderly, offering them a one-way communication interface. The methodology employs Object-Oriented Analysis and Development Methodology (OOADM) to present the system in a user-friendly manner. Additionally, an Expert System is incorporated for internal operations, mimicking human behavior by mapping knowledge into a knowledge base. The core of the system is the Speech Synthesis Module, which transforms arbitrary ASCII text into speech through the extraction of phonetic components, matching in a phonetic inventory, and generating acoustic signals for voice output. The chosen Free TTS speech engine, programmed in JAVA, supports SAPI and JSAPI, providing effective control over speech signals. This approach seeks to enhance accessibility and communication for those with physical limitations, contributing to the development of inclusive technology solutions.

[9] A Comprehensive Review-Based Study on Text-to-Speech Technologies. It aims to evaluate and compare different TTS technologies, including concatenative TTS, formant synthesis TTS, and statistical parametric TTS. The research focuses on understanding the advantages and limitations of these technologies in terms of naturalness of voice, system complexity, and suitability for different applications. Additionally, the study aims to explore the latest advancements in TTS technology, such as neural TTS and hybrid TTS. The problem statement includes the challenges faced by TTS systems, such as the lack of emotional expressiveness and naturalness in synthesized speech. The methodology involves a comprehensive review of existing TTS technologies. The study explores the working of TTS, including the types of voices (standard voice, neural voice, custom neural voice) and relevant terminology like phoneme, prosody, and mel-spectrogram. The paper provides a high-level diagram of the TTS structure, outlining components such as the preprocessor, encoder, decoder, and vocoder. The literature review section covers various studies on TTS, with a focus on Bangla language TTS systems. Different approaches, including unit selection, diphone concatenation, and rule-based systems, are discussed.

[10] Image to Text to Speech Conversion Using Machine Learning. It aims to involves the use of machine learning algorithms, particularly optical character recognition (OCR), to recognize and extract text from images. The proposed approach includes preprocessing steps such as converting images to grayscale, eliminating noise and non-text objects, and binarization. Advanced TTS technology is integrated into the system to convert the extracted text into speech. The authors mention the use of deep learning techniques for image captioning and established machine learning libraries and frameworks to implement and evaluate their models. The study also presents a block diagram illustrating the image-to-text-to-speech conversion process. the rapidly developing field of image-to-text-to-speech conversion using machine learning. The key problem is enabling the accurate extraction of text from images and converting it into speech. The authors emphasize the potential of this technology to revolutionize information interaction by combining optical character recognition (OCR) and text-to-speech (TTS) technologies. The application scenarios include making information more accessible for people with visual impairments, students, tourists, researchers, and musicians. The paper aims to contribute to improving the accuracy, efficiency, and robustness of image-to-text-to-speech conversion systems.

[11] Text To Speech Synthesizer.
In this paper involves a series of steps in the Text-To-Speech (TTS) process. Firstly, text analysis is performed by breaking down input text into words and sentences. Following this, text normalization transforms the text into a spoken form. Grapheme-to-phoneme conversion assigns phonetic transcription to each word, ensuring accurate pronunciation. The final step is speech synthesis, where a waveform corresponding to the input text is generated. The paper introduces two models, namely the Common Form model and the Grapheme-Phoneme Form model, to elucidate the text-to-speech conversion process. the problem of converting text into speech using Text-To-Speech (TTS) technology. It specifically focuses on the implementation of a computer-based system that can read any text aloud, exploring the integration of Optical Character Recognition (OCR) with speech synthesis technologies. The objective is to create a cost-effective, user-friendly image-to-speech conversion system using Python, enabling the transformation of both manually entered and scanned text into speech. The motivation behind this is the preference for listening to information rather than reading it, allowing for multitasking while consuming crucial data.

[12] An Efficient Approach for Text-to-Speech Conversion Using Machine Learning and Image Processing Technique. This novel approach aims to optimize the conversion process, potentially resulting in faster and more accurate text-to-speech synthesis. The specific methodologies within machine learning and image processing are not detailed, but the paper likely explores their synergistic use for effective and efficient text-to-speech conversion. In this paper using Two approaches, maximally stable extensible region (MSER) and grayscale conversion, are used for text character recognition. Geometric filtering combined with stroke width transform (SWT) is employed for further processing. Letter/alphabet grouping is performed to detect text sequences, fragmented into words. A 96 percent accurate spell check is carried out using naive Bayes and decision tree algorithms, The primary focus is on assisting visually impaired individuals in accessing and understanding textual content present in images, such as bus numbers, hotel names, newspapers, etc.

[13] Text to Speech Synthesis: A Systematic Review, Deep Learning Based Architecture and Future Research Direction. a comprehensive examination of the landscape of text-to-speech synthesis. Through a systematic review, the paper likely surveys existing methodologies and approaches in the field, providing a holistic understanding of the advancements and challenges. The focus then shifts to a deep learning-based architecture, indicating a novel approach grounded in deep learning techniques for improving the synthesis of speech from text. This could

involve the utilization of neural networks, potentially addressing limitations in prior methods. Moreover, the paper likely outlines future research directions, offering insights into areas where improvements and innovations can be made in the realm of text-to-speech synthesis. For a nuanced understanding of the paper's contributions, one would need to delve into specific sections such as the systematic review findings, the proposed deep learning architecture, and the outlined directions for future research. TTS technology has advanced significantly, particularly with the integration of deep learning techniques. Further research is needed to explore the identified research directions to make TTS systems even more expressive, efficient, versatile, and adaptable to diverse domains and real-time applications.

[14] Text to Speech Conversion using Raspberry – PI.
It aims Optical Character Recognition (OCR) and Text-to-Speech Synthesizer (TTS) in a Raspberry Pi. The process involves capturing an image using a camera, passing it to the TTS unit installed in the Raspberry Pi, amplifying the TTS output with an audio amplifier, and then delivering it to a speaker. The OCR technology is used for text extraction from color images, converting them into voice format. The system comprises two main modules: image processing (utilizing OCR) and voice processing (utilizing TTS). the challenge faced by visually impaired individuals in accessing text-based information. With approximately 91% of people being blind, the authors propose a low-cost solution utilizing Text-to-Speech (TTS) technology integrated with a Raspberry Pi. The system aims to convert text from images into voice or audio format, providing a means for the blind to interact with computers through a vocal interface. the successful detection of text on the image and its conversion into an audio file. The system demonstrates the capability to convert both capital and small letters, providing a comprehensive solution for text-to-speech conversion from images.

[15] Text-To-Speech Synthesis System for Kannada Language. The primary objective is likely to create a technology that converts written text in Kannada into spoken language, catering to the linguistic needs of Kannada speakers. The system's emphasis on Kannada is likely to enhance accessibility and communication for users of this specific language. The methodology likely involves language-specific considerations for phonetics and linguistic nuances, ensuring accurate and natural speech synthesis. This research contributes to the broader field of text-to-speech synthesis by addressing language-specific challenges and requirements, facilitating improved communication and information accessibility for Kannada speakers. It specifically focuses on the challenges of synthesizing natural and emotionally expressive speech in Kannada. The critical factors considered include the selection of appropriate speech units, prosody generation, and the incorporation of emotional features to enhance the naturalness of the synthesized speech. providing valuable insights into the duration, pitch, and intensity dynamics in emotional speech synthesis.

## IV. METHODOLOGY

Text-to-speech (TTS) synthesis encompasses a multi-step process for converting written text into spoken words. The initial stage involves text preprocessing, addressing punctuation and formatting issues, followed by linguistic analysis to grasp the text's structure and semantics. Prosody modeling becomes pivotal for infusing rhythm and intonation into the synthesized speech, encompassing parameters like pitch, duration, and amplitude. Additionally, considerations for emotional and expressive elements are integrated to enhance naturalness. the subsequent phase delves into phonetic analysis, where the text is scrutinized at the phonetic level to identify the pronunciation of each word. This involves the mapping of text to phonemes, the fundamental units of sound in a language, and an exploration of coarticulation effects for more authentic speech generation. Acoustic modeling plays a critical role in transforming phonetic and prosodic information into a speech signal representation. This representation can be achieved through concatenative synthesis, involving the combination of pre-recorded speech segments, or parametric synthesis, where deep learning models like recurrent neural networks or transformers generate speech waveforms based on learned parameters.

Voice synthesis, a pivotal component, may involve the utilization of a voice database containing recorded speech samples for concatenative synthesis. Alternatively, neural network models are employed for parametric synthesis. Post-processing steps may include speech enhancement techniques to refine the quality of the synthesized speech, addressing issues like noise reduction and improving overall clarity. the synthesized speech is subjected to evaluation through subjective listening tests and objective metrics gauging factors such as intelligibility and naturalness. The iterative refinement process follows, incorporating user feedback and ongoing research to continually enhance performance and achieve a more human-like quality. Notably, advancements in deep learning, particularly the utilization of sophisticated neural network architectures, have significantly propelled the quality and naturalness of text-to-speech synthesis in recent years.

## V. RESULTS

Text-to-speech synthesis aims to produce speech output that is natural, intelligible, and contextually appropriate. Ideally, the synthesized speech should closely resemble human speech patterns, capturing nuances of prosody, intonation, and rhythm to achieve a natural sound. This includes conveying emotional elements present in the input text, making the output expressive and engaging. Phonetically accurate pronunciation is crucial for clear and understandable speech, considering variations in accents and regional nuances. An effective TTS system should be adaptable to different speaking styles, contexts, and languages, providing versatility in its applications. Real-time synthesis capabilities, low latency, and high-quality audio output contribute to a seamless user experience, particularly in interactive applications and voice-controlled systems. Customization options, allowing users to modify aspects of the synthesized voice, contribute to a personalized and user-friendly experience. Despite the significant advancements driven by deep learning and neural network architectures, achieving perfect naturalness and prosody, especially in complex or emotionally expressive texts, remains an ongoing challenge that researchers and developers continue to address. The ultimate goal is to continually refine TTS technology to deliver high-quality, human-like speech synthesis across diverse linguistic contexts and applications.

## VI. CONCLUSION

Text-to-Speech (TTS) technology has had a profound impact on accessibility, communication, and user experience. TTS has revolutionized the way individuals with visual impairments or reading difficulties access written content, providing them with equal opportunities to engage with information, education, and entertainment. the advancements in TTS have led to more natural and high-quality synthesized speech, making the listening experience enjoyable and immersive. The availability of TTS in multiple languages has further broadened its reach, enabling people from diverse linguistic backgrounds to benefit from this technology. moreover, TTS systems are becoming increasingly personalized, allowing users to select voices that resonate with them or even create their own unique voices. The applications of TTS span across various industries, including assistive technologies, e-learning platforms, navigation systems, virtual assistants, audiobooks, and entertainment media. As research continues, the future of TTS holds tremendous potential for further advancements, including enhanced naturalness, expressiveness, and adaptability.